\documentclass[fleqn,usenatbib,useAMS,twocolumn]{mnras}
\usepackage{graphicx}	
\usepackage{amsmath}	
\usepackage{amssymb}	
\usepackage{multicol}        
\usepackage{bm}		
\usepackage{pdflscape}	
\usepackage[T1]{fontenc}
\usepackage{ae,aecompl}
\usepackage{newtxtext,newtxmath}
\usepackage{caption}
\usepackage{multirow}
\usepackage[utf8]{inputenc}
\usepackage{hyperref}
\usepackage{xcolor}
\usepackage{multirow}

\usepackage[T1]{fontenc}
\usepackage[english]{babel}
\usepackage[flushleft]{threeparttable} 
\usepackage{booktabs,caption}
\usepackage{blindtext}
\usepackage{subcaption}
\usepackage[flushleft]{threeparttable} 
\usepackage{booktabs,caption}
\begin{document}
\title{Probable Quasi-Periodic Oscillations in the {\tt TESS} observations of blazars in the Swift X-ray Survey }

\author[Tripathi et al.]{
Ashutosh Tripathi,$^{1,2}$\thanks{E-mail: ashutosh31tripathi@yahoo.com}
Paul J. Wiita,$^{3}$
and Krista Lynne Smith,$^{2}$
\\
$^{2}$Xinjiang Astronomical Observatory, CAS, 150 Science-1 Street, Urumqi 830011, People's Republic of China\\
$^{2}$George P. and Cynthia Woods Mitchell Institute for Fundamental Physics and Astronomy, Texas A\&M University, College Station, TX 77843-4242, USA \\
$^{3}$Department of Physics, The College of New Jersey, 2000 Pennington Rd., Ewing, New Jersey 08628-0718, USA\\
}

\pubyear{2025}

\label{firstpage}
\pagerange{\pageref{firstpage}--\pageref{lastpage}}
\maketitle
\begin{abstract}
This work presents possible quasi-periodic oscillations in the Transiting Exoplanet Survey Satellite observations of blazars that are in the 157-month hard X-ray survey done by Swift's Burst Alert Telescope. We report observations from four sources, J1104.4+3812, J1654.0+3946, J0353.4-6830, and J1941.3-6216, that show at least 3$\sigma$ local significance in generalized Lomb-Scargle periodogram and weighted wavelet Z-transform methods. These results are also checked using a continuous autoregressive moving average analysis that also predicts the absent data tentatively using stochastic differential equations. Each of these four sources exhibits a nominal QPO signal frequency in the range of 0.5--1.1 d$^{-1}$, resulting in at least 5 putative cycles. However, when the number of frequencies examined and the number of sources examined are both taken into account, the global significances are reduced to $\approx2.2\sigma$ for J1654.0+3964 (Mrk 501), and to $\approx1.9\sigma$ for other sources. These QPOs are thought to arise from the processes within the relativistic jet. Plausible explanations include the kink instability, which arises due to current-driven instabilities in the plasma or the precession of substructures, or mini-jets, within the jets.   
\end{abstract}


\section{Introduction}\label{sec:intro}

Blazars are radio-loud Active Galactic Nuclei (AGNs) with their relativistic jets pointing toward the Earth \citep{1995PASP..107..803U}. This special orientation (inclination $\theta~\lesssim~{10}^{\circ}$) leads to Doppler boosting of its emission, which usually overwhelms the brightness of the whole galaxy. They are classified into two subcategories, i.e., BL Lac and Flat Spectrum Radio Quasars (FSRQs). The emission from BL Lac objects arises almost entirely from jets, and we detect no or weak optical emission lines \citep{1991ApJ...374...72S, 1995PASP..107..803U, 1996MNRAS.281..425M, 2017MNRAS.469..255G, 2019ARA&A..57..467B}. In addition to the continuum emission from the jets of FSRQs, we see prominent broad-line emissions with equivalent width $> 5$\AA, as is typical for quasars \citep{1978PhyS...17..265B, 1998MNRAS.299..433F, 1995PASP..107..803U, 2019ARA&A..57..467B}. These radio-loud objects are found to be highly variable throughout the electromagnetic spectrum \citep[see][and references therein]{2008Natur.452..966M}. The presence of a variety of timescales over multiple wavebands provides us with a route to studying the inner regions of AGNs \citep[e.g.][and references therein]{2010Natur.463..919A,2015ApJ...807...79H,2017Natur.552..374R,2023MNRAS.526.4502R,2023ApJ...953L..28M}, and consequently the supermassive black holes, which are believed to reside at the center of these galaxies \citep{1998ApJ...509..678G}.

Quasi-periodic oscillations (QPOs) have been observed extensively in systems containing compact objects, including neutron star binaries and stellar mass black hole binaries \citep[see][for reviews]{2006ARA&A..44...49R, 2016AN....337..398M}. In neutron binaries and stellar mass black hole binaries, QPOs have been observed frequently and are believed to originate in the innermost regions of these systems. Hence, these oscillations could also probe the strong gravity regime and its effect on the surrounding material. QPOs seen in the light curves of these compact objects have been divided into two classes on the basis of frequency. The more common low-frequency QPOs (LFQPOs) lie in the frequency range of mHz to 30 Hz and could be explained by models like unstable density waves \citep{1999A&A...349.1003T} and Lens-Thirring precession \citep{2009MNRAS.397L.101I}. The relatively rarely seen high-frequency QPOs (HFQPOs) have frequencies greater than 30 Hz and are explained in the literature using models such as diskosiesmology \citep{2001ApJ...559L..25W}, disk-jet coupling \citep{2004ApJ...601..414L}, and warped accretion disks \citep{2005PASJ...57..699K}.

Unlike in stellar mass compact binaries, QPOs have rarely been convincingly observed in AGNs. However, QPOs apparently have been detected with timescales varying from a few minutes, through hours, days, weeks, and even years in all observable bands of the electromagnetic spectrum, from radio through $\gamma$-rays. For instance, periodicities of 15.7 minutes in radio observations \citep{1985Natur.314..148V} and 23 minutes in optical observations \citep{1985Natur.314..146C} from OJ 287 were detected and were explained by the presence of "hotspots" in the accretion disks. In addition, a periodicity of 11.7 years is reported in the same source using decades-long optical observations that can be explained by OJ 287 containing a binary supermassive black hole system \citep{1996A&A...305L..17S}. Quasi-periods in the range of years have also been observed in other sources in multiple wavebands \citep[see][and references therein]{2001A&A...377..396R, 2006A&A...455..871F, 2009A&A...504L...9V, 2015Natur.518...74G, 2017A&A...600A.132S, 2021MNRAS.501.5997T, 2024ApJ...977..166T}. These apparent periodicities have been explained in the literature by various scenarios, including a binary supermassive black hole, precession of the jet, and a blob or shock moving outwards through the jet.  

\citet{2022Natur.609..265J} recently claimed the periodicity of 13 hours in the optical and $\gamma$-ray observations of BL Lacertae and attributed it to the temporal growth of a plasma-driven kink in the rotating and precessing jet. Similarly, a period of 0.7 days is reported by analyzing simultaneous optical and ultraviolet observations \citep{1993ApJ...411..614U}. QPOs of the order of days has also been reported \citep[see][and references therein]{1991ApJ...372L..71Q, 1996A&A...305...42H,2003ApJ...585..665H, 2024MNRAS.527.9132T, 2024MNRAS.528.6608T}. \citet{2018ApJ...860L..10S} and \citet{2019MNRAS.484.5785G} reported quasi-periods of a few weeks, and respectively proposed to explain them using density waves and Lens-Thirring precession. These QPOs on different timescales detected in different wavebands can thus probe the physical processes occurring in various locations near the black hole or out in the jet and thus can play a crucial role in understanding these systems.  

Due to irregular, uneven, and seasonal gaps, searching for periodicity in ground-based optical observation is a great challenge, as there is always the risk of mimicking an actual periodic signal with the stochastic red noise variability present in AGN \citep[e.g][]{2016MNRAS.461.3145V}. The {\tt Kepler} mission \citep{2010Sci...327..977B} is the first space-based optical instrument that essentially overcomes these issues and provides regular, evenly sampled data that could then be used for variability studies \citep{2018ApJ...857..141S}. Using 950 days of {\tt Kepler} data from an AGN, KIC 9650712, \citet{2018ApJ...860L..10S} claimed to have found an optical QPO of 44d period. Similarly to {\tt Kepler}, its successor, the Transiting Exoplanet Survey Satellite \citep[{\tt TESS};][]{2015ESS.....350301R} provides an evenly sampled and high cadence that has been shown to be suitable for studying variability in AGN \citep{2020ApJ...900..137W, 2021MNRAS.501.1100R, 2021MNRAS.504.5629R}. Recently, \citet{2024MNRAS.527.9132T} found possible QPOs in the light curves of 2 blazars found in the Sloan Digital Sky Survey IV (SDSS-IV) catalog that were classified as BL Lacs. After a systematic search for periodicity in sources from the fourth Fermi-LAT $\gamma$-ray catalog, \citet{2024MNRAS.528.6608T} found probable oscillations in the {\tt TESS} observations of five blazars. 

In this work, we search for signatures of QPOs in {\tt TESS} observations of the blazars detected in the 157-month BAT AGN hard X-ray survey. In Sect.~ref{sec:red}, we explain the reduction procedure employed for {\tt TESS} data and the source selection process. In Sect.~\ref{sec:tech}, we describe the methods used to search for and quantify any quasi-periodic signatures, i.e., generalized Lomb-Scargle periodogram, weighted wavelet Z-transform, and continuous autoregressive moving average methods. The result for each source along with their description is given in Sect.~\ref{sec:res}. Finally, we will discuss the plausible models to explain these QPOs and the caveats in analyzing the data from {\tt TESS} in Sect.~\ref{sec:dis}. 


\begin{table*}
\centering
\caption{The sources and the corresponding {\tt TESS} observations analyzed in this work. }
\begin{tabular}{|c|c|c|c|c|c|c|c|}
\hline
Source & Common Name & {\tt TESS} Cycle (Sector) & mean flux $\pm$ $\sigma$ &  z & R mag & Type\\\hline
SWIFT J1104.4+3812 & Mrk 421 & 4 (48) & 4303.5 $\pm$ 16.1 & 0.0031& 8.31 & BL Lac\\ \hline
\multirow{3}{*}{SWIFT J1654.0+3946} & \multirow{3}{*}{Mrk 501} & 2 (24, 25)& 1671.7 $\pm$ 4.8  & \multirow{3}{*}{0.0034}&\multirow{3}{*}{17.5} & \multirow{3}{*}{BL Lac}\\
& & 4 (51, 52)& 984.7 $\pm$ 10.51 & &\\ 
& & 6 (78, 79)& 888.5 $\pm$ 5.41 & &\\ \hline
SWIFT J0353.4-6830 & PKS 0352-686 & 1 (1) & 562.8 $\pm$ 0.8  & 0.087 & 12.3 & BL Lac\\ \hline
SWIFT J1941.3-6216 & PKS 1936-623 & 1 (13) & 465.7 $\pm$ 0.8 & N/A & N/A & Quasar\\ \hline
\end{tabular}\label{tab:1}
\end{table*}

\section{Data Reduction and Source Selection}\label{sec:red}

The {\tt TESS} mission \citep{2015ESS.....350301R} is involved in long-term essentially continuous observations of stars in a search for exoplanets. {\tt TESS} eventually covered the entire sky in the optical waveband with a rapid cadence of 30 minutes in the first two cycles and as short as 200 seconds for the current one (Cycle 7). There are four cameras, each with a field of view of $24^{\circ}\times 24^{\circ}$, constituting a sector, and a combined FOV of $24^{\circ}\times 96^{\circ}$; where each pixel corresponds to $\approx$ 21 arc-seconds by 21 arc-seconds. The monitoring baselines depend on the ecliptic latitude, with lower latitudes and at poles being 27 days and $\approx$ 1 year, respectively. One sector of {\tt TESS} observations corresponds to 27 days in which an object at a low latitude can be observed before moving to the next patch of the sky. These observations are regular and evenly sampled, except for the gap in the middle of the sectors, which is in the range of 1-5 days depending on the systematics. The instrument covers one hemisphere in the course of a year and then is reoriented to scan the other hemisphere.

As {\tt TESS} is designed primarily to observe periodic signals from exoplanet passages in front of their stars, careful reduction is required to derive AGN light curves, as they contain stochastic signals which might be interpreted as systematics. {\tt Quaver} \citep{2023ApJ...958..188S} is a pipeline that uses the principal component analysis (PCA) method to remove additive background flux and multiplicative systematics. It allows one to choose the extraction aperture of a source using full-frame images (FFI) of {\tt TESS} observations using the {\tt TESSCut} package \citep{2019ASPC..523..397B}, which not only avoids contamination from nearby sources, but also makes {\tt TESS} suitable for studying extended sources. Various routines from the {\tt lightkurve} package \citep{lightkurve} are implemented to fit the background flux below a given threshold and the systematics. This information is then stored in the form of matrices and removed from the original light curve using the {\tt RegressionCorrector} routine. For details, see \citet{2023ApJ...958..188S}, \citet{2024ApJ...973...10D}, and \citet{2024MNRAS.527.9132T}. 

The {\tt TESS} observations analyzed in this work correspond to the sources found in the 157-month hard X-ray survey \citep{2025arXiv250604109L} carried out by the Burst Alert Telescope \citep[BAT;][]{2005SSRv..120..143B} onboard the Neil Gehrels Swift Observatory (henceforth, {\tt Swift}). This survey is an extension to the 105-month hard X-ray survey \citep{2018ApJS..235....4O} and consists of 1888 sources. For this work, we have selected sources that are classified there as ``beamed AGNs". Some sources in the recent survey are classified as ``possible blazar" or ``FSRQ". We included all such sources in our study sample, which constitutes a total of 206 hard X-ray sources. Some {\tt TESS} observations are affected by systematic and instrumental effects that are difficult to mitigate with confidence and thus, defy convincing analysis. Thus, these observations are dropped from the sample. 
In some cases, multiple sources are present on the same pixel, making it very difficult to isolate the data from the desired source. So, we also need to remove such sources. When this {\tt Swift} survey catalog is cross-matched with that of {\tt TESS}, 22 sources are found to be mismatched, and 8 sources have not been observed by {\tt TESS}. After these cuts, our sample consists of 126 sources which can be used for further study.

\section{Data Analysis Techniques}\label{sec:tech}
We perform periodogram analyses on all these ``good" sources to inspect whether they have a red noise power spectrum with non-zero spectral index $\alpha$, which hints at the presence of variability. In this work, the sources are considered variable if their spectral index lies in the range of 1--3 which have been reported in PSD analyses of 
observations made in different wave bands for both short and long term variability \citep{2016ApJ...824L..20A, 2018A&A...620A.185N, 2020ApJ...897...25B, 2021MNRAS.501.1100R, 2023ApJ...951...58W}. We find that 38 sources have variable red noise spectra and we further analyzed them to estimate the significance of their variability and of any putative QPOs. The detailed study of the variability of all of these sources is in progress (Tripathi et al., in preparation). 

Ten sources showed a QPO  of at least $2\sigma$ using the different analysis techniques used in this study. Finally, we find that four of the main sample of 38 sources have local significances for the claimed periods exceeding $3\sigma$ (99.73 \%), which we consider to be strong candidate QPOs, and are presented in this work. These sources are listed in Table~\ref{tab:1}. We have found possible QPOs in all three {\tt TESS} cycles of Mrk 501 and in the single sectors for which the remaining three sources were observed. The duration of each sector varies in the range of 26--29 days. All these sources have {\tt TESS} magnitudes brighter than 17.5, which is the {\tt TESS} threshold magnitude for the appropriate use of {\tt Quaver}\footnote{\url{https://github.com/kristalynnesmith/quaver}} \citep[see also][]{2024MNRAS.527.9132T}. The variation in the flux is measured to be highest for Mrk 421, followed by Mrk 501. For different observations of Mrk 501, the flux decreases over time. 

Now, we will discuss the time-series analysis methods used in this work to quantify any possible observed quasi-periodicities. These techniques are the generalized Lomb-Scargle periodogram, the weighted wavelet Z-transform, and the Continuous Auto-regressive Moving Average method. 

\subsection{Generalized Lomb-Scargle Periodogram (GLSP)}
The GLSP is a commonly used method to estimate the periodogram of a given observation and is a modified version of the Lomb-Scargle periodogram \citep{1976Ap&SS..39..447L, 1982ApJ...263..835S}. It fits the given observation with sinusoidal and constant components using likelihood optimization and also includes errors in fluxes in its computation. \citet{2009A&A...496..577Z} describes the GLSP implementation used in this work, which is also provided as a routine in {\tt PYASTRONOMY} package\footnote{\url{https://pyastronomy.readthedocs.io/en/latest/pyTimingDoc/pyPeriodDoc/gls.html}}.

\subsection{Estimating significance of a QPO}
The power spectral density (PSD) of a blazar usually follows the inverse power law where the periodogram density $P$ is inversely proportional to the frequency ($\nu$) raised to an exponent known as the spectral index $\alpha$ \citep{2014MNRAS.445..437M, 2018A&A...620A.185N, 2024ApJ...973...10D}. The value of $\alpha$ is found to be $\approx 2$, which corresponds to red noise or stochastic variability. Any quasi-periodicity, on the other hand, is believed to originate from a coherent process. To be significant, a QPO signal should stand well above the red noise spectrum in Fourier space. As the relation between PSD and frequency is inverse, any irregularities in the spectrum at lower frequencies, which have more power, could be misinterpreted as a quasi-periodic signal. In addition, there normally are PSD features arising from the limited length of any observation,  which could also produce false peaks. To mitigate the effects of these factors, it is important to simulate the red noise present in the observation and then estimate the strength of any QPO signal relative to it. 

We simulated 10,000 light curves that had the same statistical properties as the given observation and calculated the corresponding GLSP. For these simulations, we employed the {\tt Stingray} Python package\footnote{\url{https://github.com/StingraySoftware}} \citep{2019ApJ...881...39H} based on the method described in \citet{1995A&A...300..707T},  which includes not only the statistical information of an observation but also the PSD representing stochastic variability. This method assumes that the probability distribution is normally distributed, as is the case for all the sources analyzed in this work. We calculated the GLSP of all these simulated light curves and fit the power spectrum with a broken power-law as described in \citet{2024ApJ...977..166T}. For fitting, we considered the frequency range of 0.1--1.0 d$^{-1}$, which is typical of periodicities observed in previous high-cadence optical light curves \citep{2022Natur.609..265J, 2024MNRAS.527.9132T, 2024MNRAS.528.6608T}. Higher frequencies will be dominated by instrumental white noise, and at lower frequencies, there are too few measurements to properly derive a PSD slope. 
\subsubsection{Global Significance}
Through those simulations we obtained 10,000 GLSP amplitudes at each frequency. Then we estimated the p-value, or the probability of finding a signal having a strength equal to or greater than the observed QPO signal. The null hypothesis is that this signal is purely due to the red noise variability. The p-value at each frequency can be obtained by the distribution of the power density over each frequency \citep[see][for details]{2024ApJ...977..166T}. As this p-value is calculated at each frequency, it is termed a single period (or local) p-value \citep{2022ApJ...926L..35O}. The local p-value estimates the probability of finding a peak stronger than the QPO signal at the observed frequency only. The global p-value, however, estimates the probability of finding such a peak at any frequency. We followed the method described in \citet{2022ApJ...926L..35O} and list both local and global p-values for each observation in Table 2 as significance in percent ($(1-p)\times 100 \%$). For global significance, the frequency resolution is taken as 1/$\zeta T$ where $T$ is the duration of the observation and the value of $\zeta$ is decided such that the number of frequencies is equal to the number of input points in the observations \citet{1982ApJ...263..835S}, which range between 1.0 and 5.0. The lower and upper frequencies are defined as $1/T$ and $N/2T$, respectively, where $N$ is the number of observation points. After these estimations, we search for simulations exceeding the observed peak in the frequency range of 0.1--1.0 d$^{-1}$.

We searched for possible periodicity in 38 variable blazars in the BAT-AGN survey. So, in addition to the frequencies considered for computing global significance for an individual source, the number of blazars examined for periodicity should also be considered, as it is not known a priori which source(s) would exhibit periodicity \citep{2025MNRAS.541.2955P}. In this case, global p-value $p_{global}$ can be defined, in terms of local p-value $p_{local}$, as 
\begin{equation}
    p_{global} = 1 - (1 - p_{local})^Q
\end{equation}
where $Q$ is the trial factor, which is defined as  
\begin{equation}
    Q = N \times B,
\end{equation}
with $N$ and $B$ respectively being the number of independent frequencies and the number of blazars searched for periodic signatures \citep{2025MNRAS.541.2955P}. The value of $B$ is 38 in our work, but there is an ambiguity in defining $N$. \citet{2009A&A...496..577Z} estimates $N$ by the range of frequencies and frequency resolution used in the periodicity analysis. So, we follow \citet{2022ApJ...926L..35O} to construct the grid of frequencies and use the number of frequencies considered in the range of possible QPO frequencies (0.1--0.4 d$^{-1}$) found in this work \citep{2025MNRAS.541.2955P}. In this way, we obtain 16 independent frequencies for a 50-day baseline. In this case, the 4$\sigma$ and 5$\sigma$ local significances correspond to $\approx1.92\sigma$ and $\approx2.2\sigma$ global significances, respectively. As we have single-sector observations in this sample, also for which $N = 8$, the local significances of 4$\sigma$ and 5$\sigma$ yield global significance values of $\approx2.2\sigma$ and $\approx2.5\sigma$, respectively. If we evaluate the global significance for those local significance values for $N =12$, we find them to be $\approx2.0\sigma$ and $\approx2.3\sigma$. 





\subsection{Weighted Wavelet Z-transform (WWZ)}
The wavelet simultaneously decomposes the given observation in both the frequency and time domains \citep{1998BAMS...79...61T}. It not only estimates the strength of a signal, but also shows its evolution in time. The WWZ is a wavelet method used for irregular and unevenly sampled observations \citep{1996AJ....112.1709F, 2005NPGeo..12..345W}. This method uses the Morlet wavelet function to fit the observation, as compared to sinusoidal components used in GLSP. In this work, we used the publicly available AAVSO WWZ software\footnote{\url {https://www.aavso.org/software-directory}} that we have recently used to analyze {\tt TESS} data \citep{2024MNRAS.527.9132T, 2024MNRAS.528.6608T}. The marginalized power of WWZ over time gives the time-averaged WWZ, which is essentially a periodogram with two degrees of freedom \citep{1996AJ....112.1709F}, and the significance of its peak is calculated in the same way as for GLSP \citep[see][and references therein]{{2018A&A...616L...6G}, 2024ApJ...977..166T}.

\subsection{Continuous Auto-regressive Moving Average (CARMA) analysis}
The variability manifested in an AGN time series is understood to stem from its accretion disk and/or jets. The resultant variability interacts with the surrounding plasma and becomes non-linear and complex. Thus, presenting the light curves in the form of nonlinear equations could not only quantify the manifested variability but also determine its temporal evolution, which is also crucial for forecasting or interpolation of data. A CARMA (p,q) model of a process is a solution of the stochastic differential equation with autoregressive (AR) and moving average (MA) coefficients of the order of $p$ and $q$, respectively \citep{2014ApJ...788...33K,2017MNRAS.470.3027K}; $q <p$ is a necessary condition for the CARMA process to be stationary. CARMA modeling has been used extensively to study AGN variability \citep[see][and references therein]{2016A&A...585A.129S, 2017ApJ...834..111C, 2024ApJ...977..166T}. It models the light curves directly in the time domain, which has the advantage of not having spectral distortions like red-noise leak and aliasing that afflict  Fourier analyses. 

In this work, we used the CARMA (p,q) implementation provided by the {\tt EzTao}\footnote{\url{https://eztao.readthedocs.io/en/latest/}}  package as described in \citet{2022ApJ...936..132Y}. Each observation is initially fitted with the CARMA process having $p$ and $q$ in the range of $0\geq p \geq 7$ and $0\geq q \geq p$, respectively. The best model is selected on the basis of the Akaike information criterion (AIC). Once the model is selected, the light curve is modeled. The corresponding PSD is then calculated and compared with the GLSP. 


\begin{figure*}[t]
\hspace{-1cm}\includegraphics[scale=0.4]{figure_3812.png}
\caption{The {\tt TESS} light curve and time series analysis results for SWIFT J1104.4+3812 (Mrk 421). {\it Panel (a)}: {\tt TESS} observation (in black) and the corresponding CARMA fit (in orange). {\it Panel (b)}: Power spectrum calculated by generalized Lomb-Scargle periodogram (in blue) and best-fit CARMA model. {\it Panel (c)}: The wavelet color-color diagram showing the WWZ power increasing as the color goes from green to pink. {\it Panel (d)}: Significance estimation for GLSP (black) and time-averaged WWZ (red). The dashed green and blue curves correspond to 3$\sigma$ (99.73\%) significance for GLSP and WWZ, respectively. All these methods confirm the QPO of 4 days with at least 3 $\sigma$ confidence}\label{fig:3812}
\end{figure*}
\section{Results}\label{sec:res}

In this section, we discuss the results of the light curves with nominally significant QPOs obtained by employing the time-series methods discussed in the previous section. 
We have four such sources and will discuss them individually. For analysis, the light curves are binned at 0.1 days ($\sim$2.5 hours). To enforce a degree of stationarity on the light curves, which is necessary for CARMA analysis \citep{2024ApJ...977..166T}, we have removed linear trends in the data by fitting the light curve with a straight line.

\subsection{SWIFT J1104.4+3812 (Mrk 421)}
SWIFT J1104.4+3812 (J3812 onward), commonly known as Mrk 421, is the closest active galaxy in our parent sample at $z=0.031$ \citep{1999ApJ...525..176P} and was one of the first objects detected at TeV energies \citep{1992Natur.358..477P}. This BL Lac object has been studied extensively in all wavebands \citep[see][and references therein]{2020ApJS..248...29A, 2022MNRAS.513.1662M,  2025A&A...694A.195A} and is found to be extremely variable in the X-ray and $\gamma$-ray bands. QPOs of around 300 days apparently have been detected for this source in observations taken in X-rays \citep{2023ApJ...950..174S, 2018ApJ...859L..21C} and $\gamma$-rays  \citet{2023A&A...672A..86R}. \citet{2016PASP..128g4101L} also reported a $\sim$300-day QPO in the combined analysis of observations from the Fermi-LAT, Swift, and Owens Valley Radio Observatory.

\begin{figure*}
\hspace{-1cm}\includegraphics[scale=0.4]{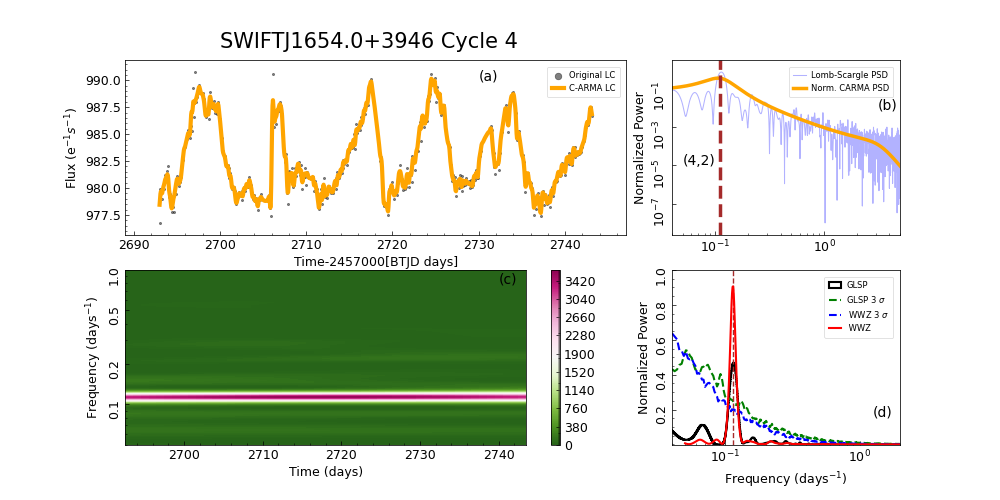}
\caption{As in Fig.\ 1 but for cycle 4 (Sec.51\& 52) {\tt TESS} observations of SWIFT J1654.0+3946 (Mrk 501). The analyses with CARMA, GLSP, and WWZ confirm the presence of a 8.7-d QPO with more than 3$\sigma$ confidence. }\label{fig:3946_4}
\end{figure*}

\begin{figure*}
\hspace{-1cm}\includegraphics[scale=0.4]{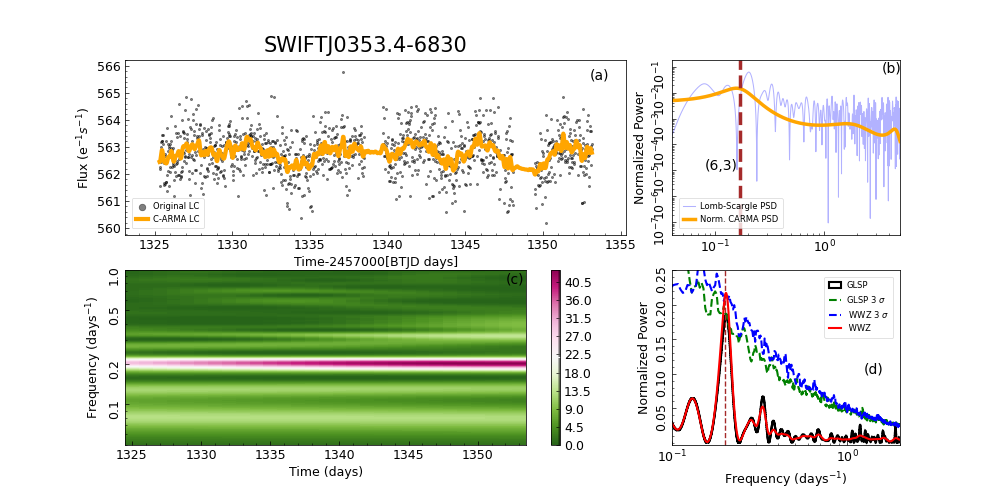}
\caption{As in Fig.\ 1 but for SWIFT J0353.4-6830 which exhibits a QPO of 5 days with at least 99.73\% significance.}\label{fig:6830}
\end{figure*}

Figure~\ref{fig:3812} shows the {\tt TESS} light curve and the corresponding results for SWIFT J1104.4+3812 (J3812 onward). The black points in panel (a) show the light curve, which shows indications of periodicity on visual inspection, and the orange curve corresponds to its fit to the CARMA (4,3) model. Panel (b) shows the Lomb-Scargle periodogram and CARMA periodogram in blue and orange, respectively, on the log-log scale.  Both periodograms show a peak at around 0.25 d$^{-1}$, which is further confirmed by the WWZ analysis presented in (c) via a color-color diagram. 
That clearly shows that most of the power is concentrated around the frequency of 0.25 d$^{-1}$ and that this signal persists throughout the observation. Panel (d) shows the significance calculations of the observed QPO peak for both GLSP and time-averaged WWZ approaches. For both periodograms, the claimed QPO peak at 0.25 d$^{-1}$ is found to be more than 3$\sigma$ significant locally. When computed globally for this observation following \citet{2022ApJ...926L..35O}, the significance of the observed peak is found to be marginally consistent with 3$\sigma$. However, when one considers the number of blazars checked for QPOs in this work, the global significance reduces to 92\%. The claimed QPO period of 4 days corresponds to 6.5 oscillations for this observation with a baseline of around 26 days. 

\subsection{SWIFT J1654.0+3946 (Mrk 501)}
SWIFT J1654.0+3946 (J1654 onward), also known as Mrk 501, is a BL Lac object situated at a redshift of 0.034 \citep{1996ApJ...456L..83Q}. This source has exhibited rapid variability in multi-wavelength observations \citep[see][and references therein]{1996ApJ...456L..83Q, 2004ApJ...600..127G, 2007ApJ...669..862A, 2024ApJ...974...50C}. Possible QPOs from this source have been reported in optical \citep{2007AcASn..48..407Y, 2023MNRAS.518.5788O} and $\gamma$-ray \citep{2019MNRAS.487.3990B}
observations.

The Cycle 4 light curve and the corresponding time series analysis results for J1654 are plotted in Figure~\ref{fig:3946_4}. This observation has two consecutive sectors (Sec.51 \& 52) with the longer baseline of 51 days. The light curve and the subsequent best fit to a CARMA (4,2) model appear to show periodic variations. The periodograms calculated by GLSP and CARMA modeling indicate the dominance  of red noise, such that the origin of variability is stochastic. Both GLSP and CARMA PSD show a peak around 0.11 d$^{-1}$ (8.7 d), which is around 6 oscillations for the observational timeline of 51 days. The WWZ color-color diagram also shows a very strong characteristic around the frequency of 0.11 d$^{-1}$ which retains nearly the same strength throughout the observation. 
The last panel plots the significance of the observed peak in GLSP and time-averaged WWZ. As evident from the local 3$\sigma$ significance curves, the claimed QPO signal is found to be very strong for both periodograms. The number of simulations performed in this work cannot distinguish between significances above 4$\sigma$, so we take the local significance to be 99.995\%. The global significance for this source alone also exceeds 4$\sigma$. Considering all the examined sources, we use Equation 1 to find an overall global significance of 97.93\%.  \\

\subsection{SWIFT J0353.4$-$6830}

SWIFT J0353.4$-$6830 (J0353 onward), commonly known as PKS 0352$-$686, is also a BL Lac object, at a redshift of 0.087 \citep{2019MNRAS.486.1741F}. This source has been studied in detail using high-energy observations, especially from Fermi-LAT \citep{2014A&A...564A...9H, 2017ApJ...850...33K, 2025arXiv250608497N}. In addition, it also has been studied in optical \citep[e.g.][]{2006A&A...455...11M} and radio \citep[e.g.][]{2014MNRAS.438..796S} wavebands. 

 The {\tt{TESS}} light curve and the corresponding CARMA (6,3) model fit in panel (a) of Fig.~\ref{fig:6830} show a possible  quasi-periodic variation in the light curve of sector 1 of this source. The CARMA PSD shows a peak around 0.18 d$^{-1}$, which is marginally consistent with that found by the GLSP method. The WWZ plot also shows a strong concentration of power around the frequency of 0.2 d$^{-1}$, which persists throughout the observation. This nominal frequency corresponds to a QPO period of 5 days, which includes almost 6 oscillations for the 28-day timeline of this observation. The QPO peak at 0.2 d$^{-1}$ is found to exceed 3$\sigma$ local significance for both GLSP and time-averaged WWZ. The global significance for this observation is found to be consistent with the 98.7\% confidence interval. But when we include the number of blazars examined, the result is not significant (55.83\%).  

\subsection{SWIFT J1941.3-6216}

SWIFT J1941.3-6216 (J1941 onward) is identified as a quasar and is commonly known as PKS 1936-623 \citep{2012MNRAS.422.1527M}. Recently, this source has been suggested to be a possible emitter of neutrinos detected by the Ice Cube Observatory \citep{2023Sci...380.1338I, 2025arXiv250116440A}.

\begin{figure*}
\hspace{-1cm}\includegraphics[scale=0.4]{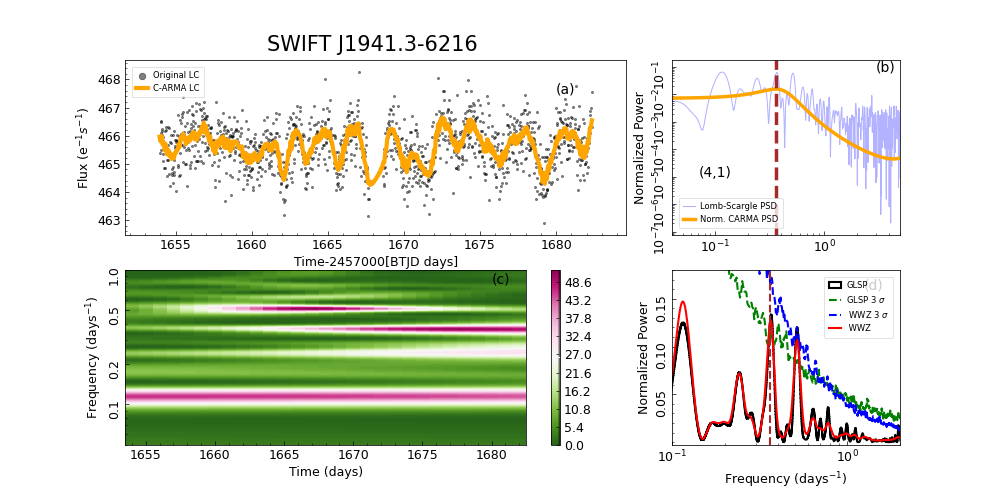}
\caption{As in Figure 1 for SWIFT J1941.3-6216. The analysis indicates the presence of 2 significant QPOs, of 1.9 and 2.7 days, which are more prominent in different portions of the observation, as shown by the WWZ color-color diagram.}\label{fig:6216}
\end{figure*}

Figure~\ref{fig:6216} first presents the sector 13 {\tt{TESS}} light curve of J1941. The orange curve is obtained by fitting the light curve with a CARMA (4,1) model. Visually, quasi-periodic features appear to be present. The corresponding PSDs are plotted in panel (b), where the CARMA PSD peaks around 0.36 d$^{-1}$, indicating a potential QPO. This putative QPO frequency is also found to be clearly present in the WWZ color-color plot in panel (c). This signal is relatively weak at the beginning of the observation, but its strength increases with time and eventually becomes dominant in the later half. There is also a signal around 0.5 d$^{-1}$ that is strongest in the middle of the observation. In panel (d), where the GLSP and the time-averaged WWZ are plotted along with their significance of 3 $\sigma$, the peak around 0.36 d$^{-1}$ is found to be more than 3$\sigma$ significant only for the GLSP, while for the time-averaged WWZ, the peak is somewhat below 3$\sigma$. The signal at 0.5 d$^{-1}$, on the other hand, is more than 99.73\% locally significant in both approaches for both periodograms. But as the peak at 0.36 d$^{-1}$ is present in all three methods, it is taken as the strongest case for a real QPO frequency for J1941 during this observation, and it corresponds to 10 oscillations. The global significance of this peak, considering the number of frequencies examined, is marginal at 95.10\%. One reason for the lower global significance is the presence of multiple peaks in the frequency interval of 0.1--1.0 d$^{-1}$. 
The peak at 0.12 d$^{-1}$ resides in the low frequency region after the break frequency, as shown in panel (b), and is probably spurious as $P\propto \nu^{-\alpha}$. 
If we remove the lower frequency peak by only considering the frequencies between 0.15 and 1.0 d$^{-1}$ for the global significance calculation, we find the significance increases to 98.05\% for the combined global significance of the peaks at 0.36 and 0.5 d$^{-1}$. However, the significance calculated using equation 1 is only 88.15\%.

\begin{table*}
 \centering\hfill
 \caption{Time series analysis results for the QPOs detected in this work. }
 \begin{threeparttable}
\begin{tabular}{|c|c|c|c|c|c|c|c|c|}
\hline
Source & Sector&$\alpha$ & QPO period (d)  & p-value (local)  & p-value (global)\tnote{a} & p-value (global)\tnote{b}&Baseline (d) & Oscillations \\\hline
J1104 & 48 & 1.64 $\pm$ 0.19 &  4.02 $\pm$ 0.06 & 99.97\% & 99.21\% & 91.94\% &26.8 & 6.7\\ \hline
\multirow{3}{*}{J1654} & 24-25 & 1.40 $\pm$ 0.13 &  10.1 $\pm$ 0.13 & 99.88\% & 99.63\% & 68.51\% &53.4 & 5.3\\ 
& 51-52 & 2.12 $\pm$0.18  & 8.7 $\pm$0.05 & 99.99\%\tnote{c} & 99.95\% & 97.93\%  & 50.1 & 5.75\\\hline
J0353 & 1 & 1.1 $\pm$ 0.12 &  4.8 $\pm$ 0.9 & 99.83\% & 98.71\% & 55.83 \% & 27.8 & 5.87\\ \hline
J1941 & 13 & 1.52 $\pm$ 0.32 & 2.7 $\pm$ 0.04 & 99.96 \%& 95.1\%\tnote{d} & 88.15\% & 28.3 & 10.48\\ \hline
\end{tabular}\label{tab:2}
\begin{tablenotes}
  \item[a] following \citet{2022ApJ...926L..35O}.
  \item[b] following Equation 1
  \item[c] The maximum significance to be obtained according to the number of artificial LCs used to estimate it.
  \item[d] See text for details. 
  \end{tablenotes}
\end{threeparttable}
\end{table*}
\begin{figure*}
\hspace{-1cm}\includegraphics[scale=0.4]{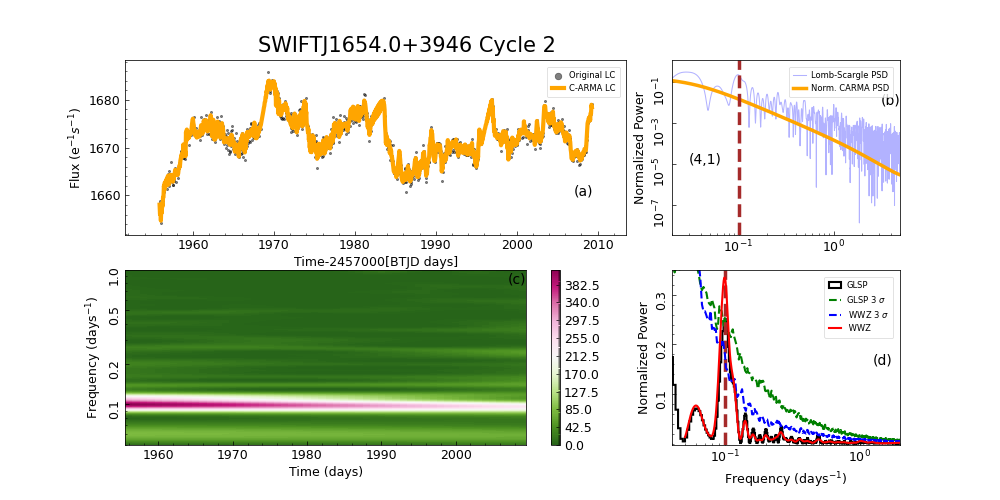}
\caption{Same as in Figure 1 but for cycle 2 observation of SWIFT J1654.0+3946. The results confirms a putative QPO of around 10 days. }\label{fig:3946_2}
\end{figure*}

\begin{figure*}
\hspace{-1cm}\includegraphics[scale=0.4]{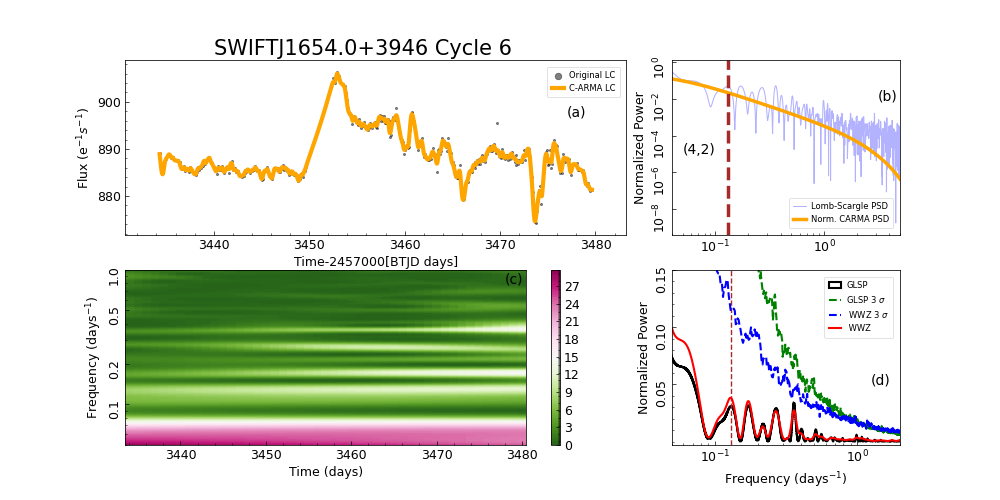}
\caption{As in Figure 1 but for the Cycle 6 observation of SWIFT J1654.0+3946. Unlike for the Cycle 2 and Cycle 4 observations of this source, this Cycle's strongest peak is around 0.13 d$^{-1}$ and is not significant.} \label{fig:3946_6}
\end{figure*}

\section{Discussion}\label{sec:dis}

This work presents a detailed quantitative analysis of the possible optical QPOs found in TESS light curves of four sources out of the 206 initial blazar members taken from the 157-month Hard X-ray Survey by the Burst Alert Telescope onboard {\tt Swift}. Our final sample consists of the 38 blazars in which variability was seen. Strong peaks are seen in the generalized Lomb-Scargle periodogram, weighted wavelet Z-transform, and continuous autoregressive moving average analyses of these four sources, which are found to have at least 99.8\% local significances, as estimated using extensive simulations, which corresponds to 3.2$\sigma$.

Table~2 lists the results of the time series analysis for the sources. The spectral index $\alpha$ is the slope of the power spectrum obtained by fitting the PSD of the observations with an inverse power-law as described in Sec.~3. We find $\alpha$ ranges from 1.0 to 2.5 (within errors) for all these sources, which is in agreement with what has been found with {\tt TESS} for most blazars \citep[e.g.][]{2020ApJ...900..137W, 2024ApJ...973...10D}.  The highly significant peaks correspond to QPO frequencies in the range of 0.11--0.52 d$^{-1}$, or periods in the range of 1.9--8.7 days in observations with timelines of 25 to 55 days. This yields the number of putative oscillations found in these observations to be in the range of 5--11,  which we consider to be enough to support the validity of the QPO claims. 

Figure~\ref{fig:3946_2} shows the results for the Cycle 2 data of J1654. A peak of around 10 days is found to be at least 3$\sigma$ significant using both GLSP and time-averaged WWZ. The WWZ power is also found to be concentrated around this frequency and persistent throughout this observation, although it weakens toward the end. However, the CARMA PSD, unlike Cycle 4, does not show any peak around this frequency. And this QPO would correspond to only around 5 oscillations.  However, we note that a plausible QPO period of around 10 days is present in two distinct observations of this source. 

These observations of J0353 belong to different Cycles, which are approximately one year apart. Detecting a QPO of similar period in two different observations strengthens the possibility of the presence of such a QPO in the intervening times when there were no {\tt{TESS}} observations. 
There is also a Cycle 6 observation available for this source for sectors 78 and 79. When analyzed collectively, as shown in Fig.~\ref{fig:3946_6}, no statistically significant periodicity is found, but a peak around 0.13 d$^{-1}$ could be seen in the GLSP. Still, this peak, at a period of around 8 days, is rather close to the QPO periods found in the other two Cycles. Thus, it could well be related to the physical process that produces significant peaks in the Cycle 2 and Cycle 4 observations and hints at the possibility of a QPO lasting over several years in Mrk 501. 

Lower global significances can be possible for the sources that show multiple peaks in the spectra. Other peaks at lower frequencies are problematic, as the red noise has larger amplitudes there \citep{2016MNRAS.461.3145V}. 
For instance, cycle 4 observation of J1654 has a prominent peak with local significance of more than 4$\sigma$. The second most prominent peak is about 10 times weaker than the QPO peak. So, the global significance that includes the other frequencies examined for this observation is also more than 3$\sigma$. However, in the PSD of J1941, there are two other peaks with power comparable to that of the nominal QPO peak and the global significance reduced to about 2$\sigma$. The significance of a given signal is also affected by noise at higher frequencies ($\geq$0.5 d$^{-1}$) as that increases the baseline of red-noise which consequently affects the shape and amplitude of power-law model used to determine significance of a given QPO peak. 

When the global significance is computed by considering not only the number of frequencies examined in each observation but also the number of sources considered in total, it is naturally reduced. As shown in Table 2, only one very strong signal remains, for a QPO of 8.7 d in the Sector 51--52 light curve of J1654 (Mrk 501).

The QPOs presented in the work correspond to periods in the range of 2--9 days. Several models could explain these timescales of the order of a few days. In our sample, J1941 is an FSRQ in which the majority of the optical band emission comes from the accretion disk. In this case, the normal modes of oscillation trapped in the localized region of the accretion disk could provide a plausible model \citep{1997ApJ...476..589P, 2004ApJ...609L..63A, 2008ApJ...679..182E}. These adiabatic oscillations are trapped in the innermost region due to the strong gravity of the central object and likely produce such QPOs. Other such models include the Lens-Thirring precession of the inner portion of the accretion disk, which naturally produces QPOs due to frame dragging \citep{1999ApJ...524L..63S, 1992A&A...259..109G}. 

The remaining three sources analyzed in this work are BL Lac objects, where almost all of the emission is believed to originate in relativistic jets \citep{2001ApJS..134..181J}. Such quasi-periodicity could also be produced inside the jet because of current-driven plasma instabilities, particularly a kink instability. These kinks are believed to arise through the interaction of the plasma instability with toroidal magnetic fields, which distorts the magnetic field and produces magnetic reconnection \citep[e/g/][]{2020MNRAS.494.1817D}. The temporal growth of the kink while traveling from the center of the jet to its boundary can produce the quasi-periodicity observed \citep{2022Natur.609..265J}; for details, see \citet{2024MNRAS.527.9132T}. This temporal growth of the kink could possibly explain the different QPO frequencies observed in multiple observation cycles of J1654. Another possibility is the presence of sub-parsec-scale geometric structures inside the jet. The periodic variations due to jet precessions are of the order of a few months to even a few years. If the periodicity is related to the size of the emission region, then it is possible that such relatively small periods could be related to the precession of possible sub-structures present inside the jets, or mini-jets \citep{2012A&A...548A.123B, 2020ApJ...903..134C}. 

If the physical processes that cause such QPOs are related to jets, then they should affect their dynamics, which will be visualized in radio images at different frequencies. 
\citet{2022Natur.609..265J} used a 43 GHz radio image of BL Lacertae to support a temporal evolution of a kink inside the jet, which is also apparently detected in the optical and $\gamma$-rays as periodic variability. These radio images at different frequencies may also unravel the geometry of the jet on a sub-parsec scale, which could also confirm the presence of mini-jets, which, as mentioned earlier, is also one of the plausible origins of these QPOs. As the mini-jets are related to the geometry of the jet and not a transient phenomenon like a kink, they could be detected more prominently in these radio observations. Further high-frequency radio observations are needed for this study. 

{\tt TESS} has proved uniquely useful in studying the variability of the order of days in AGN and providing new information on the physical process responsible for it. However, there are some issues that should be carefully evaluated. There is an inevitable gap in the middle of every {\tt TESS} observation due to the pointing of the instrument to deliver the data. This gap in data could be as long as 5 days and depends on the systematics during the pointing and on the cadence masking done during data reduction. For instance, the light curve of J1104 contains a gap of approximately 4 days, whereas there is a $\sim$1-day gap in the light curve of J1941. We have used the advanced CARMA analysis to fit the light curve with a multiple-order differential equation with several parameters and then predict the gap as shown by the orange curve in panel (a) of Figures 1-6. \citet{2024MNRAS.527.9132T} performed simulations to estimate the effects of gaps in the {\tt TESS data} on the calculation of the significance of QPO peaks and found that the errors for both cases agree with each other for frequencies above about 0.1 d$^{-1}$ but diverge below that. As all frequencies found in this work are higher than 0.1 d$^{-1}$, the gaps should have a minimal effect on our significance estimations. 

Another possible issue with {\tt TESS} observation is the short $\sim$ 27-day temporal baseline for each sector. So, any period with more than 6 days would have fewer than 4 oscillations and therefore cannot be claimed as QPO, since at least 5 oscillations are required to distinguish a stochastic process from a coherent one, causing periodicity \citep{2016MNRAS.461.3145V, 2024MNRAS.527.9132T}. Although it is quite challenging to get more than 6 oscillations with a single sector {\tt TESS} observation, it will surely provide tentative evidence of the presence of quasi-periodicity. For consecutive two-sector observations, like the Cycle 2 and 4 light curves of J1654, the claimed period is around 9 days, which also gives sufficient oscillations for the combined roughly 55-day temporal baseline. Although the Cycle 2 and Cycle 4 observations each have only a few putative oscillations, we still consider them in our analysis because a similar periodicity is found in different Cycles. In terms of QPO oscillations, the periods found in J1104 and J1941 are nominally stronger than those longer ones found in J0353 and various observations of J1654. The limited duration of these observations precludes robust and confident detection of quasi-periodicity of periods longer than a few days. If the duration of {\tt TESS} observations could be increased by lengthening the duration of sectors, then more oscillations might be detected.  We have found QPOs of similar period  in two different {\tt TESS} cycles of J1654, which are about a year apart. We also found a similar QPO signature later in Cycle 6. Even more repeated  observations could provide additional insight into the physical process responsible for such quasi-periodicities. 




 \section*{Data Availability}
The TESS data analyzed in this paper can be downloaded from the Mikulski Archive for Space Telescopes (MAST) at the Space Telescope Science Institute. 

\section*{Acknowledgement}

    We are thankful to anonymous referee for constructive comments which substantially improves the manuscript. This work was supported by the Tianshan Talent Training Program (grant No.\ 2023TSYCCX0099), the CAS `Light of West China' Program (grant No. 2021-XBQNXZ-005), and the National SKA Program of China (grant No.\ 2022SKA0120102). A.T. acknowledges the support from the Xinjiang Tianchi Talent Program. This work was also supported by the Urumqi Nanshan Astronomy and Deep Space Exploration Observation and Research Station of Xinjiang (XJYWZ2303).

{}

\end{document}